\documentclass[review]{elsarticle}

\usepackage{lineno,hyperref}
\modulolinenumbers[5]
\usepackage{amssymb}

\journal{Discrete Applied Mathematics}

\begin{document}

\begin{frontmatter}

\title{Partial order similarity based on mutual information}

\author[biofiztanszek]{Gergely Tib\'ely}
\author[kutcsop,tamop]{P\'eter Pollner}
\author[kutcsop,tamop]{Gergely Palla\corref{correspondingauthor}}
\cortext[correspondingauthor]{Corresponding author}
\ead{pallag@hal.elte.hu}
\address[biofiztanszek]{Dept. of Biological Physics, E\"otv\"os Univ., H-1117 Budapest, Hungary}
\address[kutcsop]{MTA-ELTE Statistical and Biological Physics Research Group, Hungarian Academy of Sciences, H-1117 Budapest, Hungary}
\address[tamop]{Regional Knowledge Centre, E{\"o}tv{\"os} University, H-8000 Sz{\'e}kesfeh{\'e}rv{\'a}r, Hungary}





\begin{abstract}
Comparing the ranking of candidates by different voters is an important topic in social and information science with a high relevance from the point of view of practical applications. In general, ties and pairs of incomparable candidates may occur, thus, the alternative rankings are described by partial orders. Various distance measures between partial orders have already been introduced, where zero distance is corresponding to a perfect match between a pair of partial orders, and larger values signal greater differences. Here we take a different approach and propose a similarity measure based on adjusted mutual information. In general, the similarity value of unity is corresponding to exactly matching partial orders, while a low similarity is associated to a pair of independent partial orders. The time complexity of the computation of this similarity measure is $\mathcal{O}(\left|{\mathcal C}\right|^3)$ in the worst case, and $\mathcal{O}(\left|{\mathcal C}\right|^2\ln \left|{\mathcal C}\right|)$ in the typical case of partial orders corresponding to trees with constant branching number, where $\left|{\mathcal C}\right|$ denotes the number of candidates. An interesting feature of our approach is that the similarity measure is sensitive to the position of the disagreements in the ranking: Differences at the highly ranked candidates induce  larger similarity drop compared to disagreements at the bottom candidates. 
\end{abstract}

\begin{keyword}
partial order\sep similarity measure \sep mutual information
\end{keyword}

\end{frontmatter}


\section{Introduction}
The ordering of candidates by different voters is seldom uniform in real world situations, and the related problem of rank aggregation, order similarity and voting theory in general has a very long history, dating back even to the 13th century \cite{Lullus,Borda,Condorcet}. Nevertheless, the practical applications of rank aggregation are also very important in our days and attract a wide spread scientific interest ranging from web-search \cite{Yager_web_search,Dwork_rank_agrr_web} and meta-search \cite{Aslam_meta_search,Renda_metasearch}, through information retrieval \cite{Montague_improved_retriev}, classification \cite{Cohen_learn,Lebanon_combine} to biological data bases \cite{Sese_bio_database}. The mathematical description of ranking can be given by total orders, partial orders, bucket orders, interval orders or preorders, depending on the strictness of the preferences. 

In concept, probably the most simple case is when an unambiguous preference is given for any pairs of candidates, which is naturally represented by a total order. However in practice we often encounter equivalent or incomparable pairs of candidates as well. When we allow ties between candidates, but in the mean time keep all pairs of candidates comparable, the ranking is described by a bucket order \cite{Fagin_bucket,Ailon_bucket}: The buckets are corresponding to disjoint sets of candidates, where the members of a set are all equivalent to each other and members of different sets are never in equivalence relation. To incorporate incomparable pairs of candidates, one may switch to an interval order instead: in this case an interval is associated to all candidates, and a pair of candidates is considered incomparable, if their intervals are overlapping. (Note that this is not a transitive relation, in contrast to the equivalence relation appearing in a bucket order). For candidates with non overlapping intervals, the one with the interval coming first is preferred over the other. A more general description of ranking is given by a partial order \cite{Simovici_book}, corresponding to a reflexive,  anti-symmetric and transitive relation on the set of candidates. Finally, if we do not require the relation to be anti-symmetric, we obtain a preoder. 

Evaluating the similarity between the rankings of the same candidates given by different voters is a non-trivial problem with a high importance from the point of view of practical applications \cite{Bruggemann,Klein,Monjardet,Voigt}. Accordingly, numerous different approaches have already been proposed, starting from the classical Spearman footrule distance and Kendall's tau distance for total orders \cite{Kendall_biometrika,Kendall_90,Diaconis}, through various distance measures for bucket orders \cite{Critchlow}, to the study of rank aggregation with the nearest neighbour and Hausdorff Kendall tau distances \cite{Brandenburg_12}, the comparison of partial orders via the nearest neighbour Spearman footrule distance \cite{Brandenburg_13}, and the measurement of structural dissimilarity between partial orders \cite{Fattore}. 

Here we introduce an information theoretic approach for comparing partial orders. Our method has two important novel features compared to the previously studied measures: On the one hand, we define a similarity measure instead of distance, where a perfect match is signaled by the similarity being equal to 1, and in contrast, a zero similarity value is obtained for a pair of independent rankings. On the other hand, our similarity measure is  sensitive to the position of the discrepancies in the ranking. E.g., a mismatch at the top ranks induces a larger reduction in the similarity compared to a mismatch at the bottom ranks. This property can be very useful from the point of view of practical applications: In many cases the voters care about (or know about) only the most important candidates, e.g., in case of political voting, most people are familiar only with the top politicians. Thus, when comparing the order of preference by different voters, the order of politicians at the end of the lists does not really make a difference, while discrepancies among the top names have a large impact. 

Another important application is provided by hierarchy extraction and comparing hierarchies \cite{Tibely_plosone,Caldarelli_hier_PRE}. Hierarchical organisation is an ubiquitous feature of a large variety of systems studied in natural- and social sciences. This is supported by several studies, focusing on the transcriptional regulatory network of Escherichia coli \citep{Zeng_Ecoli}, the dominant-subordinate hierarchy among crayfish \citep{Huber_crayfish}, the leader-follower network of pigeon flocks \citep{Tamas_pigeons}, the rhesus macaque kingdoms \citep{McCowan_macaque}, neural networks \citep{Kaiser_neural}, technological networks \citep{Pumain_book}, scientific journals \cite{Palgrave}, social interactions \citep{Guimera_hier_soc,our_pref_coms,Sole_hier_soc}, urban planning \citep{Krugman_urban,Batty_urban}, ecological systems \citep{Hirata_eco,Wickens_eco}, and evolution \citep{Eldrege_book,McShea_organism}. A hierarchy is usually depicted as a directed acyclic graph (DAG), where we have a single root at the top of the hierarchy, with different branches starting on the second level, leading to more sub-branches on the third level, etc. This is very similar to the Hasse diagram of a partial order, where the visualisation is capturing the ordering relations between the candidates, and it can be used to define a natural mapping between hierarchies and partial orders. There are systems where different hierarchies may be associated to the same set of entities \cite{Tibely_plosone,Palgrave}, thus, evaluating the similarity between hierarchies (or equivalently, between partial orders) can have serious importance from the point of view of practical applications.

The paper is organised as follows: in Sect.\ref{sect:partial_order_dist}. we first overview the formal definition of partial orders and the most popular previously defined distance measures between partial orders. In Sect.\ref{sect:similarity} we introduce our mutual information based approach for evaluating the similarity between partial orders. We examine the main properties of the similarity measure in Sect.\ref{sect:props}., and discuss the results in Sect.\ref{sect:disc}.

\section{Partial order distance}
\label{sect:partial_order_dist}
\subsection{Partial order}
On a domain of candidates ${\mathcal C}$ a relation ${\kappa}$ is a partial order if the following conditions hold:
\begin{itemize}
\item $\kappa$ is reflexive, i.e., $\forall x\in {\mathcal C}$ $x\preccurlyeq_{\kappa} x$,
\item $\kappa$ is anti-symmetric, i.e., $x\preccurlyeq_{\kappa} y \land x\neq y$ $\Rightarrow$ $y\not\preccurlyeq_{\kappa} x$
\item $\kappa$ is transitive, i.e., $x\preccurlyeq_{\kappa}y \land y\preccurlyeq_{\kappa}z$ $\Rightarrow$ $x\preccurlyeq_{\kappa}z$.
\end{itemize}
The intuitive interpretation of the relation $x\preccurlyeq_{\kappa}y$ is that $x$ is ranked before $y$, or $x$ is preferred over $y$. A pair of candidates are unrelated (incomparable) if $x\not\preccurlyeq_{\kappa}y \land y\not\preccurlyeq_{\kappa}x$. In case $\kappa$ becomes irreflexive, i.e., $\forall x \in {\mathcal C}$ $x\nprec_{\kappa}x$, we are speaking of a strict partial order.

 When all pairs of candidates are comparable in a strict partial order, i.e., $x\prec_{\kappa}y \lor y\prec_{\kappa}x$ $\forall x\neq y$, we obtain a total order, (which is also called as chain or a linear order in some cases). Since preference is defined for all pairs of candidates, $\kappa$ in this case defines an unambiguous order (or ranking) between the candidates, with no ties: the candidate preferred over all other candidates is ranked first, the candidate preferred over all others except the first comes second, etc. Thus, $\kappa$ in this case can be also viewed as a bijection between ${\mathcal C}$ and $\left\lbrace 1,2,\dots,\left|{\mathcal C}\right|\right\rbrace$, with $\kappa(x)$ corresponding to the position of $x$ in the ranking.

In case we have a strict partial order with incomparable pairs of candidates, and $\kappa$ is also negatively transitive beside the usual irreflexive, anti-symmetric and transitive properties, $\kappa$ can be considered as a bucket order. The negative transitivity implies that if both  $x\nprec_{\kappa}y \land y\nprec_{\kappa}x$ and  $y\nprec_{\kappa}z \land z\nprec_{\kappa}y$ then it follows that also  $x\nprec_{\kappa}z \land z\nprec_{\kappa}x$. Thus, incomparability becomes an equivalence relation in this case, and the buckets correspond to sets of candidates who are all in equivalence relation with each other in a given bucket. The negative transitivity of $\kappa$ has also the interesting consequence that if $x\prec_{\kappa}y$, then for any $z$ either $x\prec_{\kappa}z$ or $z\prec_{\kappa}y$. 

The interval orders are corresponding to partial orders with a bijection $B$ from ${\mathcal C}$ to a set of intervals as $B(x)=[l_{x},r_{x}]$ and $x\preccurlyeq_{\kappa} y$ $\Leftrightarrow$ $r_x< l_y$. The usual practice is to define the boundaries of the intervals to be integers between 1 and $\left|{\mathcal C}\right|$. Finally, in case of preorders the condition for anti-symmetry is dropped from the definition of the relation, and only reflexivity and transitivity is required.

\subsection{Distances}
\label{sect:Distances}
There are several possible alternatives for defining a distance measure between different rankings of the same candidates. For simplicity, let us start with the case of total orders. A very natural approach is given by the Kendall tau distance, corresponding to the number of inverted pairs of candidates between two total orders $\kappa$ and $\mu$, given by $K(\kappa,\mu)=\left|\lbrace \lbrace x,y \rbrace \subset {\mathcal C}: x\preccurlyeq_{\kappa}y \land y\preccurlyeq_{\mu}x \rbrace\right|$. Another classical alternative is the Spearman footrule distance, $F(\kappa,\mu)=\sum\limits_{x\in {\mathcal C}}\left|\kappa(x)-\mu(x)\right|$, where $\kappa(x)$ and $\mu(x)$ denote the position of candidate $x$ in the given rankings. A study by Diaconis and Graham has shown that the Spearman footrule distance is bounded from below by the Kendall tau distance, and by twice this value from above \cite{Diaconis}.

The problem of comparing rankings becomes more complex in case of partial orders, where ties or incomparabilities prohibit the direct use of the above mentioned simple distance measures. A possible solution is to consider a pair of total orders that do not contradict the corresponding partial orders, and take the distance between these total orders. The set of total extensions of a partial order $\kappa$, denoted by ${\rm Ext}(\kappa)$, are given by total orders such that if $\xi\in {\rm Ext}(\kappa)$, then $x\preccurlyeq_{\kappa}y \;\Rightarrow x\preccurlyeq_{\xi}y$ for all $x,y\in{\mathcal C}$. Note that the size of ${\rm Ext}(\kappa)$ can be exponential in the number of candidates. The minimum distances between the total extensions in this approach are called as the nearest neighbour Kendall tau and nearest neighbour Spearman footrule distance, given by $K_{NN}(\kappa,\mu)=\min\limits_{\xi\in{\rm Ext}(\kappa)}\min\limits_{\eta\in{\rm Ext}(\mu)} K(\xi,\eta)$ and $F_{NN}(\kappa,\mu)=\min\limits_{\xi\in{\rm Ext}(\kappa)}\min\limits_{\eta\in{\rm Ext}(\mu)} F(\xi,\eta)$ respectively.

Another possibility for measuring the distance between partial orders based on their total extensions is provided by the Hausdorff versions of the Kendall tau and Spearman footrule distances. The formal definition of these can be given as
\begin{eqnarray}
K_{H}(\kappa,\mu)&=&\max\left\lbrace \max_{\xi\in{\rm Ext}(\kappa)}\min_{\eta\in{\rm Ext}(\mu)}K(\xi,\eta),\max_{\eta\in{\rm Ext}(\mu)}\min_{\xi\in{\rm Ext}(\kappa)}K(\xi,\eta)\right\rbrace , \\
F_{H}(\kappa,\mu)&=&\max\left\lbrace \max_{\xi\in{\rm Ext}(\kappa)}\min_{\eta\in{\rm Ext}(\mu)}F(\xi,\eta),\max_{\eta\in{\rm Ext}(\mu)}\min_{\xi\in{\rm Ext}(\kappa)}F(\xi,\eta)\right\rbrace .
\end{eqnarray}

Naturally, for both the Kendall tau and the Spearman footrule distances, the nearest neighbour and the Hausdorff versions defined for partial orders coincide with the original definition of the distance measure when applied to total orders. However, the complexity of the computation of the distances can be rather different depending on what type of rankings are compared. In case of total orders, the distances can be computed in linear time. The study detailed in Ref.\cite{Brandenburg_13} has shown that the complexity of the Spearman footrule distance is linear also when comparing a pair of bucket orders, or a total order with an interval order or with a bucket order. In contrast, when at least one of the rankings to be compared is a general partial order, the computation of the Spearman footrule distance becomes an NP-complete problem.

Finally, we mention the recent approach introduced by Fattore et al. \cite{Fattore}, based on the structural dissimilarity between partial orders. Here the basic idea can be best interpreted via the Hasse diagrams of the partial orders, corresponding to DAGs, capturing the ordering relations. The nodes in such a graph represent the candidates, and there is a directed link from $x$ to $y$ if and only $x$ covers $y$, meaning that $x\preccurlyeq_{\kappa} y$, and there is no $z\in{\mathcal C}$ for which $x\preccurlyeq_{\kappa} z\preccurlyeq_{\kappa} y$. The distance introduced in Ref.\cite{Fattore} treats partial orders with isomorphic Hasse diagrams as equivalent, (being at distance zero from each other). In case the compared Hasse diagrams are not isomorphic, the distance is equal to the minimal number of link deletions and link insertions needed for transforming one of the Hasse diagrams to become isomorphic with the other one. 

\section{Partial order similarity}
\label{sect:similarity}
The concept of mutual information was originally introduced for measuring the inter dependence between a pair of random variables. Here we first summarise its most important properties, and then move on to the definition of partial order similarity measures.

\subsection{Mutual information of random variables}
\label{sect:mutinfo_set_parts}
For discrete variables $r$ and $q$ with a joint probability distribution given by $P(r_i,q_j)\equiv P(r=r_i,q=q_j)$, the mutual information is defined as 
\begin{equation}
I(r,q)\equiv \sum_{i}\sum_{j} P(r_i,q_j)\ln \left(\frac{P(r_i,q_j)}{P(r_i)P(q_j)}\right), \label{eq:mutinfo}
\end{equation} 
where $P(r_i)\equiv P(r=r_i)$ and $P(q_j)\equiv P(q=q_j)$ denote the (marginal) probability distributions of $r$ and $q$ respectively. If the two variables are independent we can write $P(r_i,q_j)=P(r_i)P(q_j)$, thus, $I(r,q)$ becomes $0$. The above quantity is very closely related to the entropy of the random variables, 
\begin{equation}
I(r,q)=H(r)+H(q)-H(r,q), \label{eq:mutinfo_entropy}
\end{equation}
where $H(r)=-\sum_iP(r_i)\ln P(r_i)$ and $H(q)=-\sum_j P(q_j)\ln P(q_j)$ correspond to the entropy of $r$ and $q$, while $H(r,q)=-\sum_{ij}P(r_i,q_j)\ln P(r_i,q_j)$ denotes the joint entropy. Based on (\ref{eq:mutinfo_entropy}), the normalised mutual information ($NMI$) can be defined as
\begin{equation}
NMI(r,q)\equiv \frac{2 I(r,q)}{H(r)+H(q)}. \label{eq:NMI_alap}
\end{equation}
This way the NMI is 1 if and only $r$ and $q$ are identical, and  0 if they are independent.

The above concept of mutual information provides a natural similarity measure for different partitions of a given set $Q=\{ q_1,q_2,\dots,q_N\}$ of disjunct subsets. Suppose $\mathbf{U}=\{ U_1,U_2,\dots,U_R\}$ and $\mathbf{V}=\{ V_1,V_2,\dots,V_T\}$ are two partitions fulfilling the following conditions: $U_i\cap U_{j}=V_k\cap V_{l}=\emptyset$ for all $i,j\in\left\{1,R\right\},\; k,l\in\left\{1,T\right\},\; i\neq j, k\neq l$,
and also $\cup_{i=1}^RU_i=\cup_{j=1}^TV_j=Q$. In this case, the probability that an element selected at random from $Q$ belongs to $U_i$ and $V_j$ is given by
\begin{equation}
P(i,j)=\frac{\left|U_i\cap V_j\right|}{\left| Q\right|},
\end{equation}
while the marginal probability $P(i)=\left| U_i\right|/\left| Q\right|$ corresponds to the probability that a randomly selected element belongs to $U_i$, and similarly $P(j)=\left|V_j\right|/\left| Q\right|$ gives the probability that a randomly selected element is in $V_j$. The mutual information between the two partitions is given by 
\begin{eqnarray}
I(\mathbf{U},\mathbf{V})&=&\sum_{i=1}^R\sum_{j=1}^TP(i,j)\ln\left(\frac{P(i,j)}{P(i)P(j)}\right)=\nonumber \\
& & \sum_{i=1}^R\sum_{j=1}^T\frac{\left|U_i\cap V_j\right|}{\left| Q\right|}\ln\left(\frac{\left|U_i\cap V_j\right|\cdot \left| Q\right|}{\left|U_i\right|\cdot \left|V_j\right|}\right).
\label{eq:MI_parts_base}
\end{eqnarray}

However, as pointed out by N.\ X.\ Vinh, J.\ Epps and J. Bailey \cite{Vinh_AMI}, one may also correct (\ref{eq:MI_parts_base}) by subtracting the expected value for a pair of random partitions. The big advantage of this approach is that the resulting similarity measure will have a zero expected value for independent random partitions. According to Ref.\cite{Vinh_AMI}, the expected value of the mutual information (\ref{eq:MI_parts_base}) for fixed subset sizes in the two partitions can be given in a closed form. I.e., if $N=\left| Q\right|$, $a_i=\left|U_i\right|$ and $b_j=\left|V_j\right|$ denote the size of $Q$ and sizes of the subsets in $\mathbf{U}$ and $\mathbf{V}$ respectively, then the expected value of the mutual information reads 
\begin{eqnarray}
\left< I(\mathbf{U},\mathbf{V})\right>&=&\sum_{i=1}^R\sum_{j=1}^T\sum_{n_{ij}=(a_i+b_j-N)^+}^{\min(a_i,b_j)}\frac{n_{ij}}{N}\ln\left(\frac{N\cdot n_{ij}}{a_ib_j}\right) \nonumber \\ & &
\times\frac{a_i!b_j!(N-a_i)!(N-b_j)!}{N!n_{ij}!(a_i-n_{ij})!(b_j-n_{ij})!(N-a_i-b_j+n_{ij})!},
\label{eq:MI_expect}
\end{eqnarray} 
where the index $n_{ij}$ denotes the number of shared elements in $U_i$ and $V_j$. Thus, the sum according to $n_{ij}$ is running from zero to $\min(a_i,b_j)$ if $a_i+b_j\leq N$, and from $a_i+b_j-N$ to $\min(a_i,b_j)$ if $a_i+b_j>N$. (In the latter case $U_i$ and $V_j$ must have at least $a_i+b_j-N$ elements in common). Therefore, the $(a_i+b_j-N)^+$ in the expression above is equal to zero if $a_i+b_j\leq N$, and is simply $a_i+b_j-N$ if $a_i+b_j>N$. 

Based on (\ref{eq:MI_expect}), the adjusted mutual information between $\mathbf{U}$ and $\mathbf{V}$ was defined in Ref.\cite{Vinh_AMI} as 
\begin{equation}
AMI(\mathbf{U},\mathbf{V})=\frac{I(\mathbf{U},\mathbf{V})-\left< I(\mathbf{U},\mathbf{V})\right>}{\frac{1}{2}\left(H(\mathbf{U})+H(\mathbf{V})\right)-\left< I(\mathbf{U},\mathbf{V})\right>}.
\label{eq:AMI_base}
\end{equation}
This measure is 1 if and only if when $\mathbf{U}$ and $\mathbf{V}$ are identical, and its expected value is 0 for independent random partitions. 

\subsection{Mutual information of partial orders}
\label{sect:mut_info_preorders}
In the following we shall apply the concept of normalised mutual information to the problem of partial order similarity. The basic idea is to map the problem of comparing relations $\kappa$ and $\mu$ onto the problem of comparing partitions of candidates. This mapping enables the use of information theoretic measures for deciding to what extent are the compared relations similar to each other. We assume that both $\kappa$ and $\mu$ are defined over the same domain of candidates $\mathcal{C}$. Thus, for each candidate $x$ we can define the down sets according to the two relations as
\begin{eqnarray}
D_{\kappa}(x)&=&\{ y\in\mathcal{C}: x\preccurlyeq_{\kappa}y \}, \\
D_{\mu}(x)&=&\{ y\in\mathcal{C}:x\preccurlyeq_{\mu} y\},
\end{eqnarray}
containing the candidates $x$ is preceding according to $\kappa$ and according to $\mu$, respectively. The list of the down sets provides an alternative description of a partial order, e.g., the rankings of the candidates can be fully reconstructed from the down sets. Thus, an intuitive approach for comparing ranking is to examine the overlaps between the down sets imposed by the rankings. Along this line, a similarity measure closely related to the mutual information was already introduced in \cite{Tibely_plosone}. However, the quantity defined there cannot be regarded as mutual information in the strict mathematical sense, for reasons such as e.g., the missing joint probability distribution and marginal distributions between the random variables. 

In order to give well grounded definition, a natural idea would be to formulate a joint distribution between two random variables based on the sizes of the set intersections $\left|D_{\kappa}(x)\cap D_{\mu}(y)\right|$, and then use the mutual information calculated from the joint distribution and the marginal distributions for evaluating the similarity between $\kappa$ and $\mu$. However, due to the necessary normalisation of the joint distribution, the similarity measure obtained in this way will have a couple of counter intuitive properties, as described in the Appendix. 

To avoid such difficulties, here we first introduce two indicator variables $i_{\kappa}(x)$ and $j_{\mu}(x)$, associated to each candidate in the rankings $\kappa$  and $\mu$ as follows. By choosing a candidate $y$ from $\mathcal{C}$ at random we require
\begin{equation}
i_{\kappa}(x)=\left\lbrace\begin{array}{ll}
1 & \mbox{ if } y\in D_{\kappa}(x), \\
0 & \mbox{ otherwise,}
\end{array}\right.  \;\;\;
j_{\mu}(x)=\left\lbrace\begin{array}{ll}
1 & \mbox{ if } y\in D_{\mu}(x), \\
0 & \mbox{ otherwise}.
\end{array}\right.
\end{equation}
Thus, $i_{\kappa}(x)$ and $j_{\mu}(x)$ are indicating whether a candidate $y$ picked at random from $\mathcal{C}$ is in the down set of $x$ according to $\kappa$ or according to $\mu$, respectively. (A similar idea was presented in the problem of comparing overlapping community partitions of networks in Ref.\cite{Santo_mutinfo}). The joint probability distribution of $i_{\kappa}(x)$ and $j_{\mu}(x)$ can be given as
\begin{eqnarray}
P(i_{\kappa}(x)=1,j_{\mu}(x)=1)&=&\frac{\left| D_{\kappa}(x)\cap D_{\mu}(x)\right|}{\left| \mathcal{C}\right|}, \label{eq:joint_dist_1}\\
P(i_{\kappa}(x)=1,j_{\mu}(x)=0)&=&\frac{\left|D_{\kappa}(x)\right|}{\left|\mathcal{C}\right|}-\frac{\left| D_{\kappa}(x)\cap D_{\mu}(x)\right|}{\left| \mathcal{C}\right|}, \\
P(i_{\kappa}(x)=0,j_{\mu}(x)=1)&=&\frac{\left|D_{\mu}(x)\right|}{\left|\mathcal{C}\right|}-\frac{\left| D_{\kappa}(x)\cap D_{\mu}(x)\right|}{\left| \mathcal{C}\right|}, \\
P(i_{\kappa}(x)=0,j_{\mu}(x)=0)&=&1-\frac{\left|D_{\kappa}(x)\right|}{\left|\mathcal{C}\right|}-\frac{\left|D_{\mu}(x)\right|}{\left|\mathcal{C}\right|}+\frac{\left| D_{\kappa}(x)\cap D_{\mu}(x)\right|}{\left| \mathcal{C}\right|}.
\end{eqnarray}
The marginals are simply given as 
\begin{eqnarray}
& & P(i_{\kappa}(x)=1)=\frac{\left|D_{\kappa}(x)\right|}{\left|\mathcal{C}\right|} , \;\;\;
P(i_{\kappa}(x)=0)=1-\frac{\left|D_{\kappa}(x)\right|}{\left|\mathcal{C}\right|}, \\
& & P(j_{\mu}(x)=1)=\frac{\left|D_{\mu}(x)\right|}{\left|\mathcal{C}\right|},   \;\;\;
P(j_{\mu}(x)=0)=1-\frac{\left|D_{\mu}(x)\right|}{\left|\mathcal{C}\right|}.
\end{eqnarray}
Based on the above, the mutual information between $i_{\kappa}(x)$ and $j_{\mu}(x)$ can be given as 
\begin{equation}
I(i_{\kappa}(x),j_{\mu}(x))=\sum_{i_{\kappa}(x)=0}^{1}\sum_{j_{\mu}(x)=0}^1P(i_{\kappa}(x),j_{\mu}(x))\ln\left(\frac{P(i_{\kappa}(x),j_{\mu}(x))}{P(i_{\kappa}(x))P(j_{\mu}(x))}\right), \label{eq:I_alap}
\end{equation}
while the entropies of $i_{\kappa}(x)$ and $j_{\mu}(x)$ can be expressed as
\begin{eqnarray}
H(i_{\kappa}(x))&=&\sum_{i_{\kappa}(x)=0}^1P(i_{\kappa}(x))\ln P(i_{\kappa}(x)), \\
 H(j_{\mu}(x))&=&\sum_{j_{\mu}(x)=0}^1 P(j_{\mu}(x))\ln P(j_{\mu}(x)).
\end{eqnarray}

A natural way for defining the the mutual information between $\kappa$ and $\mu$ is to sum over the candidates as
\begin{eqnarray}
I(\kappa,\mu)&=&\sum_{x\in\mathcal{C}}I(i_{\kappa}(x),j_{\mu}(x))=\nonumber \\
& &\sum_{x\in\mathcal{C}}\sum_{i_{\kappa}(x)=0}^{1}\sum_{j_{\mu}(x)=0}^1P(i_{\kappa}(x),j_{\mu}(x))\ln\left(\frac{P(i_{\kappa}(x),j_{\mu}(x))}{P(i_{\kappa}(x))P(j_{\mu}(x))}\right).
\label{eq:I_def}
\end{eqnarray}
In order to be able to normalise this quantity, we can define the entropies of the partial orders in a similar fashion as
\begin{eqnarray}
H(\kappa)&=&\sum_{x\in\mathcal{C}}H(i_{\kappa}(x))=\sum_{x\in\mathcal{C}}\sum_{i_{\kappa}(x)=0}^1P(i_{\kappa}(x))\ln P(i_{\kappa}(x)), \label{eq:H_kap_def}\\
H(\mu)&=&\sum_{x\in\mathcal{C}}H(j_{\mu}(x))=\sum_{x\in\mathcal{C}}\sum_{j_{\mu}(x)=0}^1 P(j_{\mu}(x))\ln P(j_{\mu}(x)). \label{eq:H_mu_def}
\end{eqnarray}
By combining the expressions above with (\ref{eq:I_def}) we can write the normalised mutual information between $\kappa$ and $\mu$ as
\begin{equation}
NMI(\kappa,\mu)=\frac{I(\kappa,\mu)}{\frac{1}{2}(H(\kappa)+H(\mu))}.
\label{eq:NMI_def}
\end{equation}
This quantity is equal to one if and only $\kappa$ and $\mu$ are identical, and is expected to yield a small value in case $\kappa$ and $\mu$ are independent partial orders. Therefore, (\ref{eq:NMI_def}) is providing a natural candidate for a similarity measure between partial orders.

\subsection{Adjusted mutual information based on a combinatorial null model}

An intuitive requirement towards a similarity measure is to provide a zero value when a pair of random partial orders are compared, or at least to have a zero expected value for random partial orders, analogously to the adjusted mutual information between set partitions discussed in Sect.\ref{sect:mutinfo_set_parts}. We can achieve this simply by correcting the $NMI$ given in (\ref{eq:NMI_def}) by the expected value of the $NMI$ over all possible partial orders, in a similar fashion to (\ref{eq:AMI_base}) introduced for set partitions in \cite{Vinh_AMI}. However, there are several different possibilities for defining the overall set of possible partial orders. Here first for simplicity we assume that the sample space for a partial order $\kappa$ is corresponding to all possible permutations of the candidates among the different ``positions'' according to $\kappa$, i.e., the ``structure'' of the partial order is left unchanged, only the candidates are swapped randomly. (An illustration is given in Fig.\ref{fig:zero_inv}a). A more general sample space is discussed in Sect.\ref{sect:rewire}., where we also allow the ``rewiring'' of the ``structure'' of the partial order.

According to the assumption above, the total number of samples for a partial order over a set of candidates $\mathcal{C}$ is simply $\left|\mathcal{C}\right|!$. The expected value for the mutual information of independent random permutations of the candidates in $\kappa$ and in $\mu$ can be expressed as
\begin{equation}
\left< I(\kappa,\mu)\right>=\frac{1}{\left|\mathcal{C}\right|!}\frac{1}{\left|\mathcal{C}\right|!}\sum_{\pi_{\kappa}}\sum_{\pi_{\mu}}\sum_{x\in\mathcal{C}}I(i_{\kappa,\pi_{\kappa}}(x),j_{\mu,\pi_{\mu}}(x)),
\label{eq:I_expect_alap}
\end{equation}
where $\pi_{\kappa}$ and $\pi_{\mu}$ denote the permutation of the candidates in $\kappa$ and in $\mu$ respectively, and $i_{\kappa,\pi_{\kappa}}(x)$ and $j_{\mu,\pi_{\mu}}(x)$ stand for the indicator variables of the set of successors of $x$ according to $\kappa$ and according to $\mu$ when taking the position dictated by $\pi_{\kappa}$ and $\pi_{\mu}$. However, since all permutations of the candidates among the positions in a given partial order are statistically equivalent to each other,  (\ref{eq:I_expect_alap}) can be also given as
\begin{eqnarray}
\left< I(\kappa,\mu)\right>&=&\frac{1}{\left|\mathcal{C}\right|!}\sum_{\pi_{\mu}}\sum_{x\in\mathcal{C}}I(i_{\kappa}(x),j_{\mu,\pi_{\mu}}(x))= \label{eq:I_expect_kap}\\
& &\frac{1}{\left|\mathcal{C}\right|!}\sum_{\pi_{\kappa}}\sum_{x\in\mathcal{C}}I(i_{\kappa,\pi_{\kappa}}(x),j_{\mu}(x)),\label{eq:I_expect_mu}
\end{eqnarray}
where $\kappa$ can correspond to any particular permutation of the candidates from $\pi_{\kappa}$ in (\ref{eq:I_expect_kap}), and similarly $\mu$ can be any particular permutation of the candidates from $\pi_{\mu}$ in (\ref{eq:I_expect_mu}). We continue by reordering the sum in (\ref{eq:I_expect_kap}) as 
\begin{equation}
\left< I(\kappa,\mu)\right>=\frac{1}{\left|\mathcal{C}\right|!}\sum_{x\in\mathcal{C}}\sum_{\pi_{\mu}}I(i_{\kappa}(x),j_{\mu,\pi_{\mu}}(x)), 
\end{equation}
 thereby fixing $D_{\kappa}(x)$, (corresponding to the down set of $x$ according to $\kappa$) when carrying out the sum according to $\pi_{\mu}$.  By grouping the different permutations $\pi_{\mu}$ according to the position of $x$ in $\mu(\pi_{\mu})$, which we denote by $y$, we can write 
\begin{equation}
\left< I(\kappa,\mu)\right>=\frac{1}{\left|\mathcal{C}\right|!}\sum_{x\in\mathcal{C}}\sum_{y\in\mathcal{C}}\sum_{\pi_{\mu}:x\rightarrow y}I(i_{\kappa}(x),j_{\mu,\pi_{\mu}}(x\rightarrow y)),
\label{eq:I_M_is_also_fixed}
\end{equation}
where the summation according to $\pi_{\mu}:x\rightarrow y$ is running over permutations $\pi_{\mu}$ where the position of $x$ is given by $y$, and the corresponding indicator variable is denoted by $j_{\mu,\pi_{\mu}}(x\rightarrow y)$. The advantage of this grouping of the terms is that the size of the set $D_{\mu,\pi_{\mu}}(y)$, (containing the down set of $y$ according to $\mu$ at the position given by $\pi_{\mu}$) becomes fixed in the sum over $\pi_{\mu}:x\rightarrow y$. Let us denote the number of common candidates between $D_{\kappa}(x)$ and $D_{\mu}(y)$ as $c_{xy}$. According to (\ref{eq:I_alap}) the mutual information $I(i_{\kappa}(x),j_{\mu,\pi_{\mu}}(x\rightarrow y))$ can be fully evaluated based on the set sizes $\left|D_{\kappa}(x)\right|$, $\left|D_{\mu}(y)\right|$ and the size of the intersection $c_{xy}=\left|D_{\kappa}(x)\cap D_{\mu}(y)\right|$. Therefore, the terms in the sum over $\pi_{\mu}:x\rightarrow y$ in (\ref{eq:I_M_is_also_fixed}) can be grouped according to $c_{xy}$, providing 
\begin{equation}
\left< I(\kappa,\mu)\right>=\frac{1}{\left|\mathcal{C}\right|!}\sum_{x\in\mathcal{C}}\sum_{y\in\mathcal{C}}\sum_{c_{xy}} N(x,y,c_{xy})\cdot I(x,y,c_{xy}),
\label{eq:I_multiplicity}
\end{equation}
where $N(x,y,c_{xy})$ denotes the number of permutations $\pi_{\mu:x\rightarrow y}$ where $D_{\kappa}(x)$ and $D_{\mu}(y)$ have $c_{xy}$ candidates in common, and $I(x,y,c_{xy})$ can be given as
\begin{eqnarray}
I(x,y,c_{xy})&=&\sum_{i_x=0}^1\sum_{j_y=0}^1 P(i_x,j_y,c_{xy})\ln\left( 
\frac{P(i_x,i_y,c_{xy})}{P(i_x,c_{xy})P(j_y,c_{xy})}\right)=\nonumber \\
& &\frac{c_{xy}}{\left|\mathcal{C}\right|}\ln\left(\frac{c_{xy}\cdot\left|\mathcal{C}\right|}{\left|D_{\kappa}(x)\right|\cdot
\left|D_{\mu}(y)\right|}\right)+\nonumber \\
& &
\frac{\left| D_{\kappa}(x)\right|-c_{xy}}{\left|\mathcal{C}\right|}
\ln\left(\frac{\left(\left| D_{\kappa}(x)\right|-c_{xy}\right)
\left|\mathcal{C}\right|}{\left|D_{\kappa}(x)\right|
\left(\left|\mathcal{C}\right|-\left|D_{\mu}(y)\right|\right)}\right)
+ \nonumber \\
& &
\frac{\left| D_{\mu}(y)\right|-c_{xy}}{\left|\mathcal{C}\right|}
\ln\left(\frac{\left(\left| D_{\mu}(y)\right|-c_{xy}\right)
\left|\mathcal{C}\right|}
{\left(\left|\mathcal{C}\right|-\left|D_{\kappa}(x)\right|\right)\left|D_{\mu}(y)\right|}\right)+\nonumber \\
& &
\frac{\left|\mathcal{C}\right|-\left|D_{\kappa}(x)\right|
-\left|D_{\mu}(y)\right|+c_{xy}}{\left|\mathcal{C}\right|}\times \nonumber \\
& &
\ln\left(\frac{\left(\left|\mathcal{C}\right|
-\left|D_{\kappa}(x)\right|-\left|D_{\mu}(y)\right|+c_{xy}\right)
\left|\mathcal{C}\right|}
{\left(\left|\mathcal{C}\right|-\left|D_{\kappa}(x)\right|\right)
\left(\left|\mathcal{C}\right|-\left|D_{\mu}(y)\right|\right)}\right).
\label{eq:Ixycxy}
\end{eqnarray}

In order to complete the calculation of $\left<I(\kappa,\mu)\right>$ the term $N(x,y,c_{xy})$ has to be also evaluated in (\ref{eq:I_multiplicity}), which can be carried out as follows.  First we choose the common elements $c_{xy}$ from $D_{\kappa}(x)$, yielding ${\left|D_{\kappa}(x)\right|}\choose{c_{xy}}$. Next, the remaining elements in $D_{\mu}(y)$ have to be chosen from the elements not in $\left| D_{\kappa}(x)\right|$, and we cannot choose $x$ either, thus, we have a further ${\left|\mathcal{C}\right|-1-\left| D_{\kappa}(x)\right|}\choose{\left|D_{\mu}(y)\right|-c_{xy}}$ factor. The candidates in $D_{\mu}(y)$ can be at any position, bringing in a $\left|D_{\mu}(y)\right|!$ factor, and also, the candidates outside $D_{\mu}(y)$ can also be at any position, (except for $x$ placed at $y$), yielding a further $\left(\left|\mathcal{C}\right|-1-\left|D_{\mu}(y)\right|\right)!$ factor. Taken together we arrive to
\begin{eqnarray}
& &N(x,y,n_{xy})=
{{\left|D_{\kappa}(x)\right|}\choose{c_{xy}}} {{\left|\mathcal{C}\right|-1-\left| D_{\kappa}(x)\right|}\choose{\left|D_{\mu}(y)\right|-c_{xy}}} \left|D_{\mu}(y)\right|!\left(\left|\mathcal{C}\right|-1-\left|D_{\mu}(y)\right|\right)!= \nonumber \\
& &\frac{\left|D_{\kappa}(x)\right|!\left|D_{\mu}(y)\right|!\left(\left|\mathcal{C}\right|-1-\left|D_{\kappa}(x)\right|\right)!\left(\left|\mathcal{C}\right|-1-\left|D_{\mu}(y)\right|\right)!}{c_{xy}!\left(\left|D_{\kappa}(x)\right|-c_{xy}\right)!
\left(\left|D_{\mu}(y)\right|-c_{xy}\right)!\left(\left|\mathcal{C}\right|-1-\left|D_{\kappa}(x)\right|-\left|D_{\mu}(y)\right|+c_{xy}\right)!},
\label{eq:I_expect_final}
\end{eqnarray}
which is analogous to the second factor in (\ref{eq:MI_expect}). Thus, the expected value of the mutual information between $\kappa$ and $\mu$ can be written as
\begin{eqnarray}
& &\left< I(\kappa,\mu)\right>=\frac{1}{\left|\mathcal{C}\right|!}\sum_{x\in\mathcal{C}}\sum_{y\in\mathcal{C}}\sum_{c_{xy}} I(x,y,c_{xy})\times \nonumber \\ & &
\frac{\left|D_{\kappa}(x)\right|!\left|D_{\mu}(y)\right|!\left(\left|\mathcal{C}\right|-1-\left|D_{\kappa}(x)\right|\right)!\left(\left|\mathcal{C}\right|-1-\left|D_{\mu}(y)\right|\right)!}{c_{xy}!\left(\left|D_{\kappa}(x)\right|-c_{xy}\right)!
\left(\left|D_{\mu}(y)\right|-c_{xy}\right)!\left(\left|\mathcal{C}\right|-1-\left|D_{\kappa}(x)\right|-\left|D_{\mu}(y)\right|+c_{xy}\right)!}, \nonumber \\
& &
\end{eqnarray}
where $I(x,y,c_{xy})$ is given by (\ref{eq:Ixycxy}). 

Based on the above, our proposed similarity measure is corresponding to the adjusted mutual information ($AMI$) between $\kappa$ and $\mu$, which can be expressed as
\begin{equation}
AMI(\kappa,\mu)=\frac{I(\kappa,\mu)-\left<I(\kappa,\mu)\right>}{\frac{1}{2}(H(\kappa)+H(\mu))-\left<I(\kappa,\mu)\right>},
\label{eq:AMI_part_order}
\end{equation}
where $I(\kappa,\mu)$ is defined in (\ref{eq:I_def}) and the entropies of the partial orders are given in (\ref{eq:H_kap_def}-\ref{eq:H_mu_def}).

\subsection{Adjusted mutual information based on empirical corrections}
\label{sect:rewire}
In a more general framework we can also allow the occurrence of partial orders with different structure in the sample space of random partial orders. When comparing a pair of random partial orders from this space, opposed to a simple difference in the permutation of the candidates in the different positions, we may also observe differences in number of down sets and the sizes of the  down sets. However, by allowing this much larger variety of random partial orders, the tracking of the expected value of the mutual information between a randomly chosen pair of partial orders becomes analytically unfeasible. Nevertheless, from a practical point of view we may still try to evaluate this expected value empirically by sampling from the allowed set of partial orders.  

In order to allow an easy to implement sampling, we define the space of allowed partial orders using the Hasse diagram of the partial order. As already mentioned in Sect.\ref{sect:Distances}., the Hasse diagram corresponds to a DAG, in which there is a link from candidate $x$ pointing to candidate $y$ if and only $x$ is covering $y$. (This means that $x\preccurlyeq_{\kappa} y$ and $\not\exists z\in\mathcal{C}: z\neq x, z\neq y, x\preccurlyeq_{\kappa} z \land z\preccurlyeq_{\kappa} y$). Due to the anti-symmetric property of the partial order, directed loops cannot occur in the obtained graph, thus, it is acyclic. The partial order itself can be very easily reconstructed from this DAG, e.g., the down set of a candidate is equivalent to the out component of the corresponding node, (given by the sub graph that can be reached from the node from the node following the out links). 


The sample space we use for random partial orders is corresponding to all possible single rooted DAGs with a fixed number of nodes and fixed number of links. Therefore, in this section we restrict our studies to the comparison between partial orders over the same set of candidates $\mathcal{C}$ and having the same number of links in the DAG representation given above. The expected value for the mutual information between a randomly chosen pair of partial orders is calculated by sampling, i.e., constructing random DAGs with the given number of nodes and links. The corresponding adjusted mutual information between $\kappa$ and $\mu$ is given by
\begin{equation}
EMI(\kappa,\mu)=\frac{I(\kappa,\mu)-\left<I(\kappa,\mu)\right>_{\rm emp}}{\frac{1}{2}\left(H(\kappa)+H(\mu)\right)-\left<I(\kappa,\mu)\right>_{\rm emp}},
\label{eq:EMI}
\end{equation}
where $\left<I(\kappa,\mu)\right>_{\rm emp}$ denotes the obtained empirical average for the mutual information between random partial orders having the same number of links in the DAG representation as $\kappa$ and $\mu$. (In order to clearly separate the adjusted mutual information based on the combinatorial null model, the adjusted mutual information based on empirical corrections is denoted by $EMI$).

\section{Properties of the partial order similarities}
\label{sect:props}

We have defined three closely related quantities for evaluating the similarity between partial orders, given by the normalised mutual information, $NMI$, in (\ref{eq:NMI_def}), the adjusted mutual information in a combinatorial null model, $AMI$, in (\ref{eq:AMI_part_order}), and the adjusted mutual information based on empirical corrections, $EMI$, in (\ref{eq:EMI}). All of these measures are equal to 1 if and only when the compared partial orders are identical. Furthermore, the expected value of the $AMI$ and the $EMI$ is 0 for random independent partial orders. Next, we are going to estimate the time complexity of the evaluation of the AMI in Sect.\ref{sect:complexity}. This is followed by the analysis of the sensitivity of the measures to the position of the disagreements in the compared partial orders in Sect.\ref{sect:sensitivity}. 

\subsection{Complexity of the evaluation of the $AMI$}
\label{sect:complexity}
The mutual information $I(\kappa,\mu)$ can be evaluated in linear time as the function of the number of comparable pairs of candidates. In the worst case, when we are dealing with a total order, this is scaling as $\left| {\mathcal C}\right|^2$. However, a typical partial order is having a considerably smaller number of comparable pairs. For simplicity let us consider partial orders corresponding to regular trees, where the candidates in a given branch are all preceded by the candidate from which the branch was started. Under such conditions the number of comparable pairs in the partial order can be estimated as $\left|{\mathcal C}\right|\ln\left|{\mathcal C}\right|$. In any case, the time complexity is dominated by the complexity of the evaluation of $\left< I(\kappa,\mu)\right>$, as demonstrated below.

The $I(x,y,c_{xy})$ and the $N(x,y,c_{xy})$ terms can be calculated in constant time, thus, the number of operations needed for evaluating $\left< I(\kappa,\mu)\right>$ is determined by the total number of terms in the sums $\sum_x\sum_y\sum_{c_{xy}}$ appearing in (\ref{eq:I_expect_final}). The innermost sum goes from 0 (or more) to $\min(\left|D_{\kappa}(x)\right|,\left|D_{\mu}(y)\right|)$, hence the number of terms altogether in $\sum_y\sum_{c_{xy}}$ can be approximated from above as $\sum_y\left|D_{\mu}(y)\right|$, corresponding to the total number of comparable pairs of candidates in $\mu$. As mentioned above, the number of comparable pairs is scaling as $\left|{\mathcal C}\right|^2$ in the worst case, however for a partial order corresponding to a regular tree, it is scaling only as $\left|{\mathcal C}\right|\ln\left|{\mathcal C}\right|$. 

Finally, the outermost sum $\sum_x$ multiplies the number of terms by a factor of $\left|{\mathcal C}\right|$. Thus, the time complexity of the evaluation of $S$ is ${\mathcal O}(\left|{\mathcal C}\right|^3)$ in the worst case, when we are comparing two total orders. However, in the typical case of partial orders corresponding to regular trees, the complexity is ${\mathcal O}(\left|{\mathcal C}\right|^2\ln \left|{\mathcal C}\right|)$.

\subsection{Sensitivity to the position of the disorders}
\label{sect:sensitivity}

An interesting property of the defined similarity measures is that they are sensitive to the position of the mismatch between the partial orders: disagreements at the top ranks cause larger drop in the similarity compared to disagreements in the candidates ranked behind. The intuitive reason for this is that if $x$ is preceding $y$, then the size of the set corresponding to $x$ is larger than the size of the set corresponding to $y$ in our mapping, i.e., if $x\preccurlyeq_{\kappa} y$, then $\left|D_{\kappa}(x)\right|>\left|D_{\kappa}(y)\right|$. Since the similarity measures are depending on set intersections and set sizes, changes in the larger sets is expected to affect the similarity more compared to changes in the smaller sets. 

\begin{figure}[hbt]
\centerline{\includegraphics[width=0.75\textwidth]{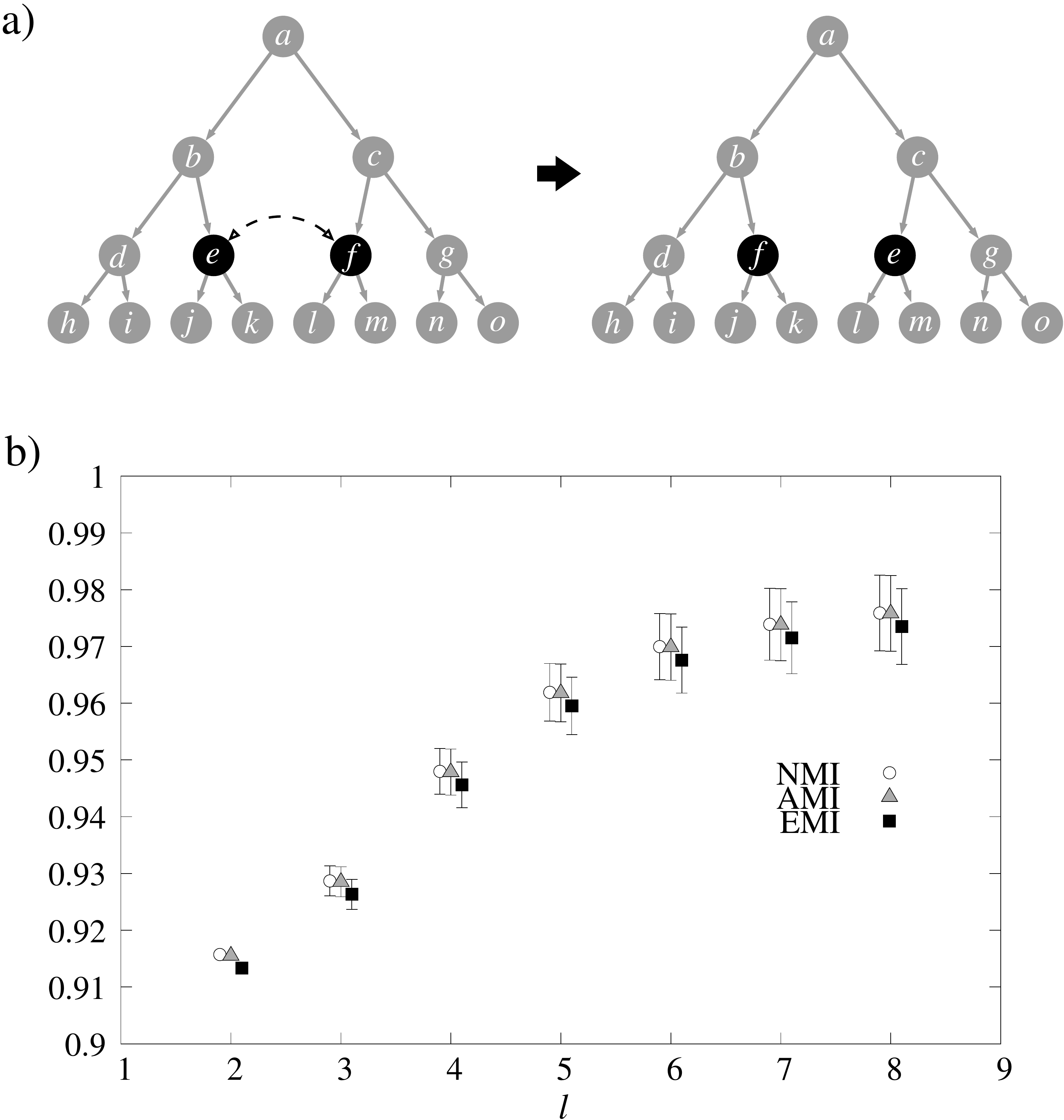}}
\caption{Comparing a partial order corresponding to a binary tree of $255$ nodes to the partial order obtained by swapping a pair of candidates on the same level $l$. a) Illustration of the partial order before and after the swap between the two candidates. Note that the Kendall tau distance is zero between the partial orders. b) The $NMI$ defined in (\ref{eq:NMI_def}), the $AMI$ given in (\ref{eq:AMI_part_order}) and the $EMI$ written in (\ref{eq:EMI}) between the partial orders as a function of the level depth $l$ of the swapped candidates, where the root is considered to be on level $l=1$. The errorbars are corresponding to the standard variation. (The symbols for the $NMI$ and for the $EMI$ have been slightly shifted horizontally for better visibility). }
\label{fig:zero_inv}
\end{figure}

We illustrate the property explained above by a simple experiment on partial orders corresponding to binary trees in the DAG representation described in Sect.\ref{sect:rewire}. According to that, the candidates in a given branch are all preceded by the candidate from which the branch was started. By swapping a pair of candidates on the same level in a given binary tree we obtain a different partial order, (for an illustration see Fig.\ref{fig:zero_inv}a). Note that since the candidates on the same level are incomparable, the Kendall tau distance between these two partial orders is zero. In contrast, the similarities are sensitive to such changes, as shown in Fig.\ref{fig:zero_inv}b, where we plot the measured $NMI$, $AMI$ and $EMI$ between the two partial orders as a function of the level depth $l$ of the swapped candidates. According to the results, when swapping candidates at the top levels, the corresponding similarities are relatively low, and show an increasing tendency as a function of $l$. 
 This is a clear signature of the sensitivity to the position of the disagreement between the compared partial orders: rearranging the top ranks of the partial order is inducing a drastic drop in the similarity, while changes in the last ranks are accompanied only by a milder decay.

\begin{figure}[hbt]
\centerline{\includegraphics[width=0.75\textwidth]{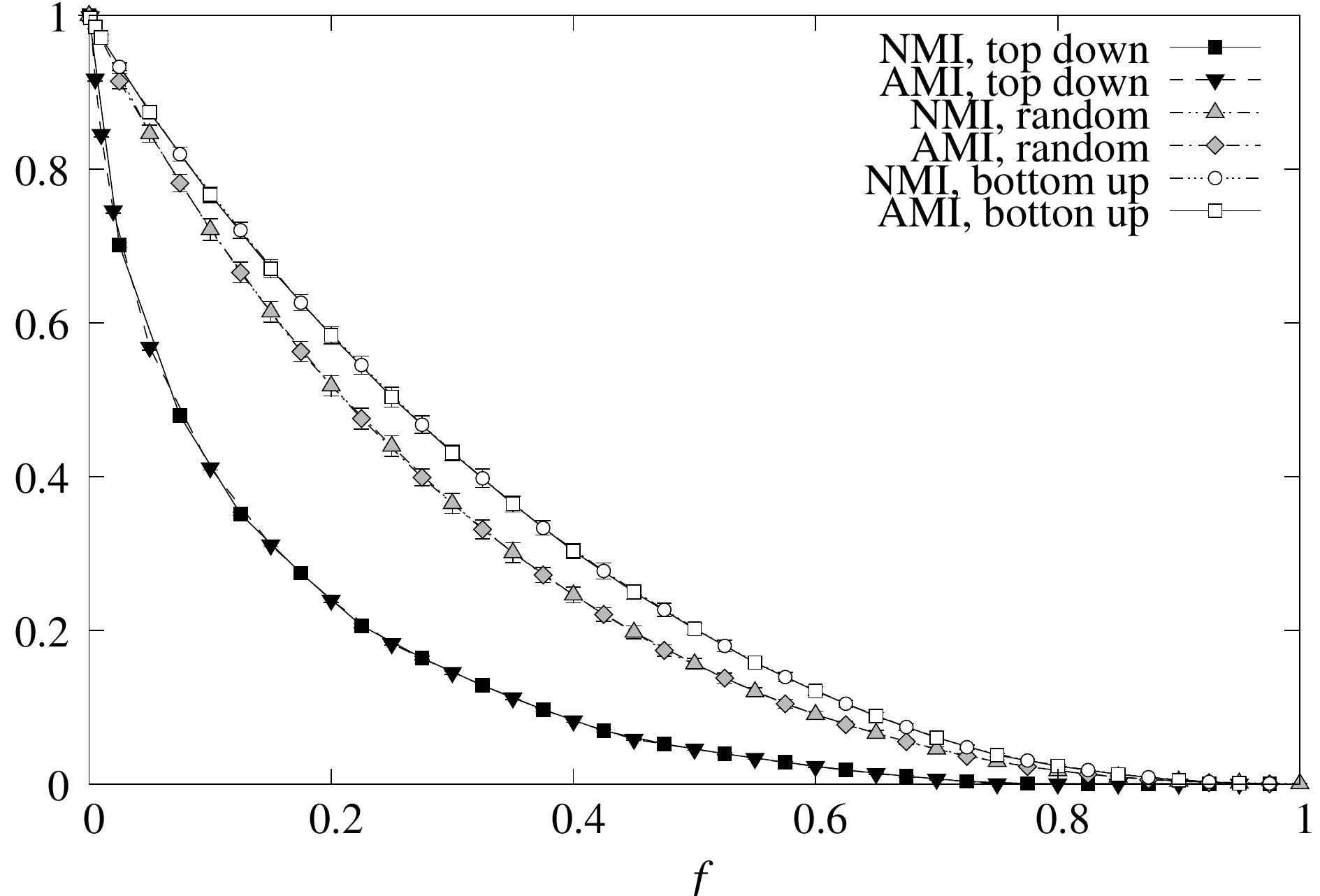}}
\caption{The $NMI$ given in (\ref{eq:NMI_def}) and the $AMI$ defined in (\ref{eq:AMI_part_order}) between a partial order corresponding to a binary tree of $2047$ nodes and its randomised counterpart as a function of the fraction of the rearranged candidates. The three different curves show the decay in the similarity measures for three different randomisation schemes: starting at the bottom ranks according to the partial order (empty symbols), rearranging the candidates in random order (grey symbols), and starting at the top ranks according to the partial order (filled symbols). Note that the error bars are mostly hidden by data points.}
\label{fig:sim_decay_plot}
\end{figure}

In Fig.\ref{fig:sim_decay_plot}.\ we show the results obtained for the $NMI$ and $AMI$ when instead of simply swapping a single pair of candidates, the partial order is completely randomised, allowing also changes in the hierarchy level of the candidates in the DAG representation. (During the randomisation procedure a randomisation step was corresponding to the swap between a pair of candidates in the partial order.) Naturally, as we increase the fraction of randomised candidates, $f$, in the partial order, the similarity  between the actual state and the original initial state is decreasing, eventually reaching zero at $f=1$. However, according to Fig.\ref{fig:sim_decay_plot}., the similarities are decreasing faster if the candidates to be swapped are chosen absolutely at random (grey symbols) compared to the case where we start the randomisation in the reverse order of the candidates (empty symbols), and an even steeper decay can be observed when we randomise according to the order of the candidates (filled symbols). 
The significant differences between the curves corresponding to the different types of randomisations is in agreement with the findings related to Fig.\ref{fig:zero_inv}. and provides a further demonstration of the sensitivity of the similarity measures to the position of the disorder in the partial order. (The behaviour of the Kendall tau distance during the randomisation is discussed in the Appendix.) Meanwhile it is also clear from the figure that the $NMI$ and the $AMI$ are practically equivalent to each other in this example, as the curves for the two types of similarities are very close to each other.

\begin{figure}[hbt]
\centerline{\includegraphics[width=0.75\textwidth]{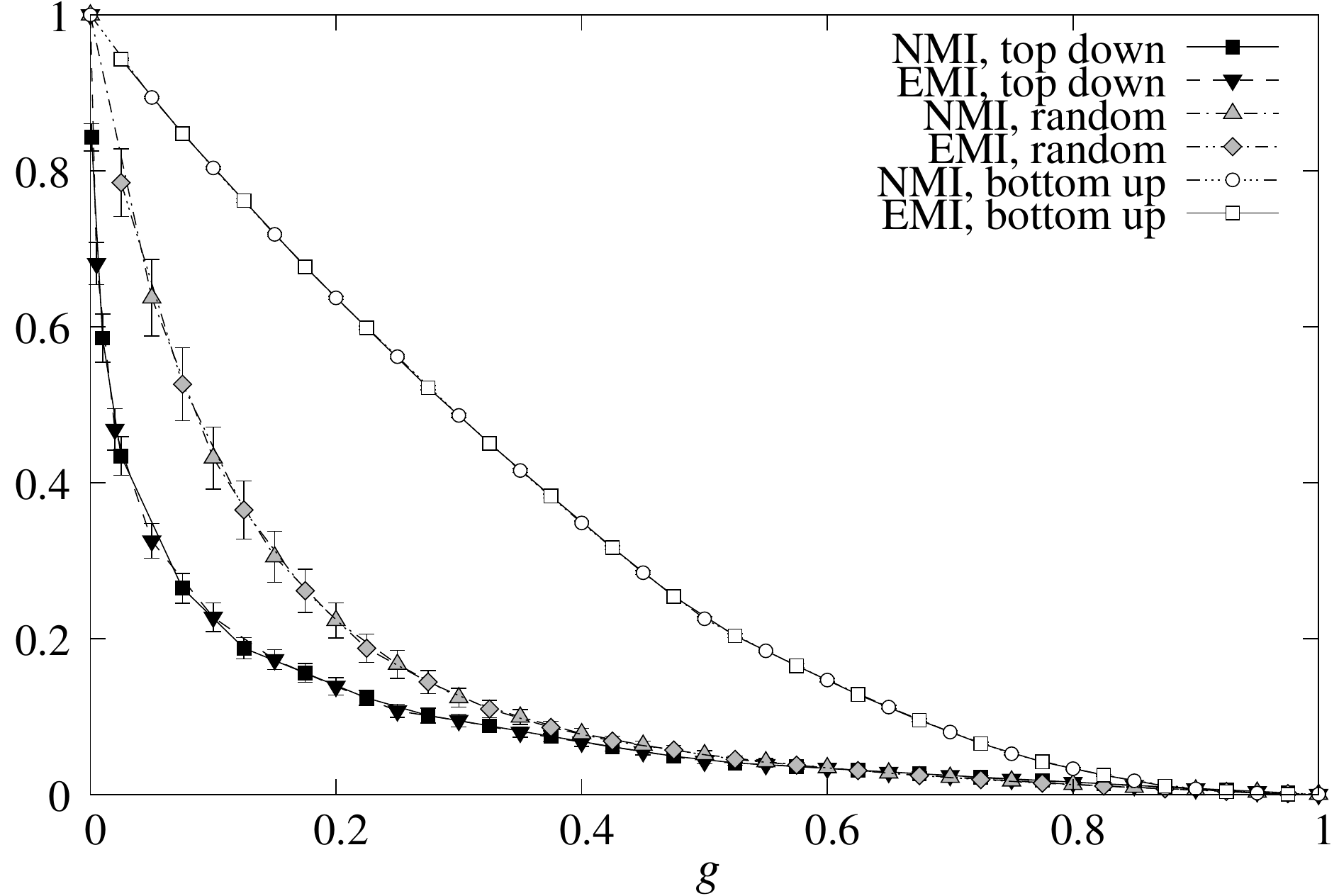}}
\caption{The $NMI$ given in (\ref{eq:NMI_def}) and the $EMI$ defined in (\ref{eq:EMI}) between a partial order corresponding to a binary tree of $2047$ nodes and its randomised counterpart as a function of the fraction of the rewired links $g$ in the DAG representing the partial order. Similarly to Fig.\ref{fig:sim_decay_plot}., the three different curves show the decay in the similarity measures for three different randomisation schemes: starting at the bottom ranks according to the partial order (empty symbols), rewiring the links in random order (grey symbols), and starting at the top ranks according to the partial order (filled symbols). Note that the error bars are mostly hidden by data points.}
\label{fig:sim_decay_rewire}
\end{figure}

Finally, in Fig.\ref{fig:sim_decay_rewire}. we show the results obtained when the randomisation is corresponding to the rewiring of the graph structure in the DAG representation of the partial order. In this case a randomisation step is corresponding to the random relocation of a link in the DAG, without breaking the graph into disconnected components. Similarly to Fig.\ref{fig:sim_decay_plot}., the curves for the $NMI$ and for the $EMI$ are very close to each other. Meanwhile, a very clear difference can be obtained between the results obtained for the three different randomisation schemes: When starting the rewiring with connections close to the root of the DAG, (filled symbols), the similarities drop very fast as a function of the fraction of the rewired links, $g$. A somewhat slower decay can be observed when the links to be rewired are chosen at random (grey symbols) and the slope of the curves is significantly smaller when rewiring the links ``bottom up'' (empty symbols). 

\section{Discussion}
\label{sect:disc}
We have introduced partial order similarity measures based on an information theoretical approach. The basic idea behind our method is to map the partial order comparison problem onto a set partition comparison problem, and apply the concept of normalised mutual information for comparing the partitions. The resulting similarity measure is equal to one if and only if the compared partial orders are identical, and is providing a value close to zero for independent random partial orders. By suitable adjustment we can switch from the NMI to similarity measures having a zero expected value for independent random partial orders. Here we have examined two possibilities, differing in the choice for the allowed space of random partial orders. 

When using a simple combinatorial null-model, the expected value for the NMI can be calculated analytically, thus, the resulting adjusted mutual information can be given in a closed form. In this case the randomisation of a partial order is corresponding to random swaps between the candidates in the different ``positions'' of the partial order, without any change in the structure of the partial order. (E.g., the topology in the DAG representation is left intact, only the node indices are permuted). The time complexity of the evaluation of the resulting AMI is ${\mathcal O}(\left| {\mathcal C}\right|^3)$ in the worst case, when we are comparing total orders. However, when comparing partial orders corresponding to regular trees, the complexity of the method is ${\mathcal O}(\left|{\mathcal C}\right|^2\ln \left|{\mathcal C}\right|)$.

We studied a more general version of the adjusted mutual information as well, where the space of allowed random partial orders is defined via the DAG representation, corresponding to all possible single rooted DAGs with a fixed number of nodes and links. Under these settings, the randomisation of the partial order is corresponding to random rewiring of the links in the DAG. Thus, instead of simply permuting the candidates, here we are allowed to also change the structure of the partial order during a randomisation process. In this framework, the expected value of the NMI can be calculated numerically averaging over a suitably large set of random samples.  

An interesting feature of the similarity measures is that they are sensitive to the position of the disagreements. I.e., we have shown in simple experiments that if the compared partial orders match at the top ranks and disagree only in the lower ranks, their similarity is likely to be higher than in the opposite case, where the disagreements occur at the highly ranked candidates. A natural interpretation of this property is that from the point of view of the similarities, the top ranks in the partial order are more important than the bottom ranks. This is consistent with the general experience that most people in every day life also tend to concentrate only on the top ranks of a wide variety of rankings occurring in sports, politics, etc. The intuitive explanation for this behaviour of the similarity measure is that in the mapping between the partial order and the set partition problem, the sizes of the sets associated to the top candidates are larger compared to the sets associated to the bottom candidates. Since the mutual information is depending on the intersections and sizes of these sets, changes in the top ranks are expected to induce changes in the larger sets and thereby affect the proposed similarities more compared to changes in the bottom ranks.

\section*{Acknowledgement}
We are very grateful to Ulrik Brandes for bringing our attention to the problem of partial order comparison, and for the extremely valuable discussions and support during the research. The work was partially supported by the European Union and the European Social Fund through project FuturICT.hu (grant no.:TAMOP-4.2.2.C-11/1/KONV-2012-0013) and by the Hungarian National Science Fund (OTKA K105447).

\section*{Appendix A}
\renewcommand{\theequation}{A\arabic{equation}}
\setcounter{equation}{0}
\renewcommand{\thefigure}{A\arabic{figure}}
\setcounter{figure}{0}

When trying to define a preorder similarity directly based on the intersection sizes of the down sets, the obtained measure is having a couple of counter intuitive properties. The main idea here is to start from the number of candidates in the intersection between the down set of candidate $i$ according to $\kappa$ and the down set of candidate $j$ according to $\mu$, expressed as
\begin{equation}
n_{ij}=\left| D_{\kappa}(i)\cap D_{\mu}(j)\right|.
\end{equation}
In order to define a joint probability distribution between two variables based on $n_{ij}$, we need to normalise $n_{ij}$ as
\begin{equation}
P(i,j)=\frac{n_{ij}}{\sum_{i,j\in\mathcal{C}}n_{ij}}=\frac{\left| D_{\kappa}(i)\cap D_{\mu}(j)\right|}{\sum_{i,j\in \mathcal{C}}\left| D_{\kappa}(i)\cap D_{\mu}(j)\right|}.
\end{equation}
The marginal distributions can be given as
\begin{eqnarray}
P(i)&=&\sum_{j\in\mathcal{C}}P(i,j)=\frac{\sum_{j\in\mathcal{C}}\left| D_{\kappa}(i)\cap D_{\mu}(j)\right|}{\sum_{i,j\in \mathcal{C}}\left| D_{\kappa}(i)\cap D_{\mu}(j)\right|}, \\
P(j)&=&\sum_{i\in\mathcal{C}}P(i,j)=\frac{\sum_{i\in\mathcal{C}}\left| D_{\kappa}(i)\cap D_{\mu}(j)\right|}{\sum_{i,j\in \mathcal{C}}\left| D_{\kappa}(i)\cap D_{\mu}(j)\right|}.
\end{eqnarray}
Thus, in principle we can define a mutual information between the two variables as,
\begin{eqnarray}
I(i,j)&=&\sum_{i,j\in\mathcal{C}}P(i,j)\ln\left(\frac{P(i,j)}{P(i)P(j)}\right)=\nonumber \\ & &
\sum_{i,j\in\mathcal{C}}\frac{\left| D_{\kappa}(i)\cap D_{\mu}(j)\right|}{\sum_{i,j\in \mathcal{C}}\left| D_{\kappa}(i)\cap D_{\mu}(j)\right|}\ln\left(\frac{\left| D_{\kappa}(i)\cap D_{\mu}(j)\right|\sum\limits_{i,j\in \mathcal{C}}\left| D_{\kappa}(i)\cap D_{\mu}(j)\right|}{\sum\limits_{j\in\mathcal{C}}\left| D_{\kappa}(i)\cap D_{\mu}(j)\right|\sum\limits_{i\in\mathcal{C}}\left| D_{\kappa}(i)\cap D_{\mu}(j)\right|}\right), \label{eq:I_ulr}
\end{eqnarray}
and the corresponding entropies as
\begin{eqnarray}
H(i)&=&\sum_{i\in\mathcal{C}}P(i)\ln P(i)=\nonumber \\
& &\sum_{i\in\mathcal{C}}\frac{\sum_{j\in\mathcal{C}}\left| D_{\kappa}(i)\cap D_{\mu}(j)\right|}{\sum_{i,j\in \mathcal{C}}\left| D_{\kappa}(i)\cap D_{\mu}(j)\right|}\ln \frac{\sum_{j\in\mathcal{C}}\left| D_{\kappa}(i)\cap D_{\mu}(j)\right|}{\sum_{i,j\in \mathcal{C}}\left| D_{\kappa}(i)\cap D_{\mu}(j)\right|}, \label{eq:H_ulr_i}\\
H(j)&=&\sum_{j\in\mathcal{C}}P(j)\ln P(j)=\nonumber \\
& &\sum_{j\in\mathcal{C}}\frac{\sum_{i\in\mathcal{C}}\left| D_{\kappa}(i)\cap D_{\mu}(j)\right|}{\sum_{i,j\in \mathcal{C}}\left| D_{\kappa}(i)\cap D_{\mu}(j)\right|}\ln \frac{\sum_{i\in\mathcal{C}}\left| D_{\kappa}(i)\cap D_{\mu}(j)\right|}{\sum_{i,j\in \mathcal{C}}\left| D_{\kappa}(i)\cap D_{\mu}(j)\right|}. \label{eq:H_ulr_j}
\end{eqnarray}

However, a rather counter intuitive property of the formalism given above is that both (\ref{eq:H_ulr_i}) and (\ref{eq:H_ulr_j}) depend on both $\kappa$ and $\mu$ simultaneously. In contrast, in case of the indicator variable approach proposed in Sect.\ref{sect:mut_info_preorders}., the entropy $H(i_{\kappa}(x))$ in (\ref{eq:H_kap_def}) is depending solely on $\kappa$, while the entropy $H(j_{\mu}(x))$ is depending solely on $\mu$. Thus, here the random variables cannot be associated to a single preorder out of the two preorders to be compared, instead they depend on a mixture of the two preorders at the same time. The other counter intuitive property of the mutual information defined in (\ref{eq:I_ulr}) is that it cannot be normalised as simply as in case of (\ref{eq:NMI_def}). I.e., if we calculate 
\begin{equation}
NMI(i,j)=\frac{I(i,j)}{\frac{1}{2}(H(i)+H(j))},
\end{equation}
in general its value is not equal to one even when $\kappa$ and $\mu$ are identical. The lack of a proper normalisation for the similarity measure can undermine its usability from the practical point of view. Based on the above, we choose the formally more elaborate, but well behaving method of defining the mutual information via indicator variables, as discussed in Sect.\ref{sect:mut_info_preorders}.
  
\section*{Appendix B}

\renewcommand{\theequation}{B\arabic{equation}}
\setcounter{equation}{0}
\renewcommand{\thefigure}{B\arabic{figure}}
\setcounter{figure}{0}

\begin{figure}[hbt]
\centerline{\includegraphics[width=0.7\textwidth]{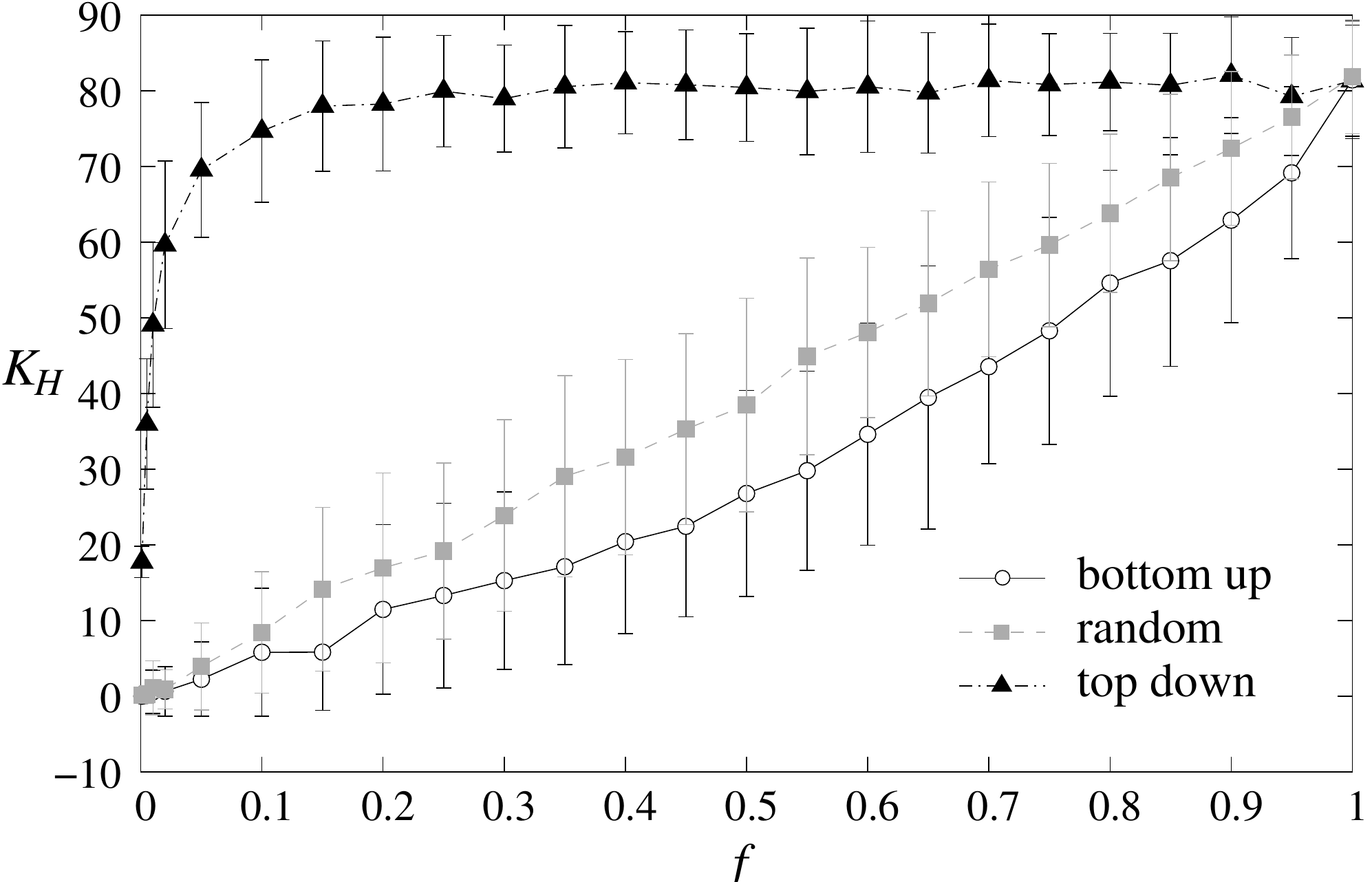}}
\caption{
The Kendall tau distance in the same experiment as shown in Fig.\ref{fig:sim_decay_plot}. The data points are corresponding to the average number of inversions between a partial order corresponding to a binary tree of $2047$ nodes and its randomised counterpart as a function of the fraction of the rearranged candidates. Similarly to Fig.\ref{fig:sim_decay_plot}., the three different curves show the results for three different randomisation schemes: starting at the bottom ranks according to the partial order (empty circles), rearranging the candidates in random order (grey squares), and starting at the top ranks according to the partial order (filled triangles).}
\label{fig:Kendal_tau}
\end{figure}

In Fig.\ref{fig:sim_decay_plot}. in the main text we have seen that the similarity $S$ defined in (\ref{eq:AMI_part_order}) is decaying as a function of the fraction of randomised candidates when randomising a partial order corresponding to a binary tree of candidates. For comparison, in Fig.\ref{fig:Kendal_tau}.\ we show the Kendal tau distance, (corresponding to the number of inversions between the compared partial orders) in the same experiment. Apparently the distance between the original partial order and its randomised counterpart is increasing as a function of the fraction of the rearranged candidates, $f$. Furthermore, the difference between the three randomisation schemes can be also observed in the behaviour of $K_H$, as rearrangement at the top ranks of the partial order is accompanied by faster increase in the average distance.    
However, note that the variance of the curves is far larger compared to Fig.\ref{fig:sim_decay_plot}. E.g., a significant part of the ``bottom-up'' data points are within the error bars of the ``random'' data points, thus, a clear distinction between these two randomisation schemes cannot be made here. Therefore, the similarity measure defined in (\ref{eq:AMI_part_order}) provides a somewhat more precise tool for comparing partial orders than the Kendall tau distance.

\begin{figure}[hbt]
\centerline{\includegraphics[width=0.9\textwidth]{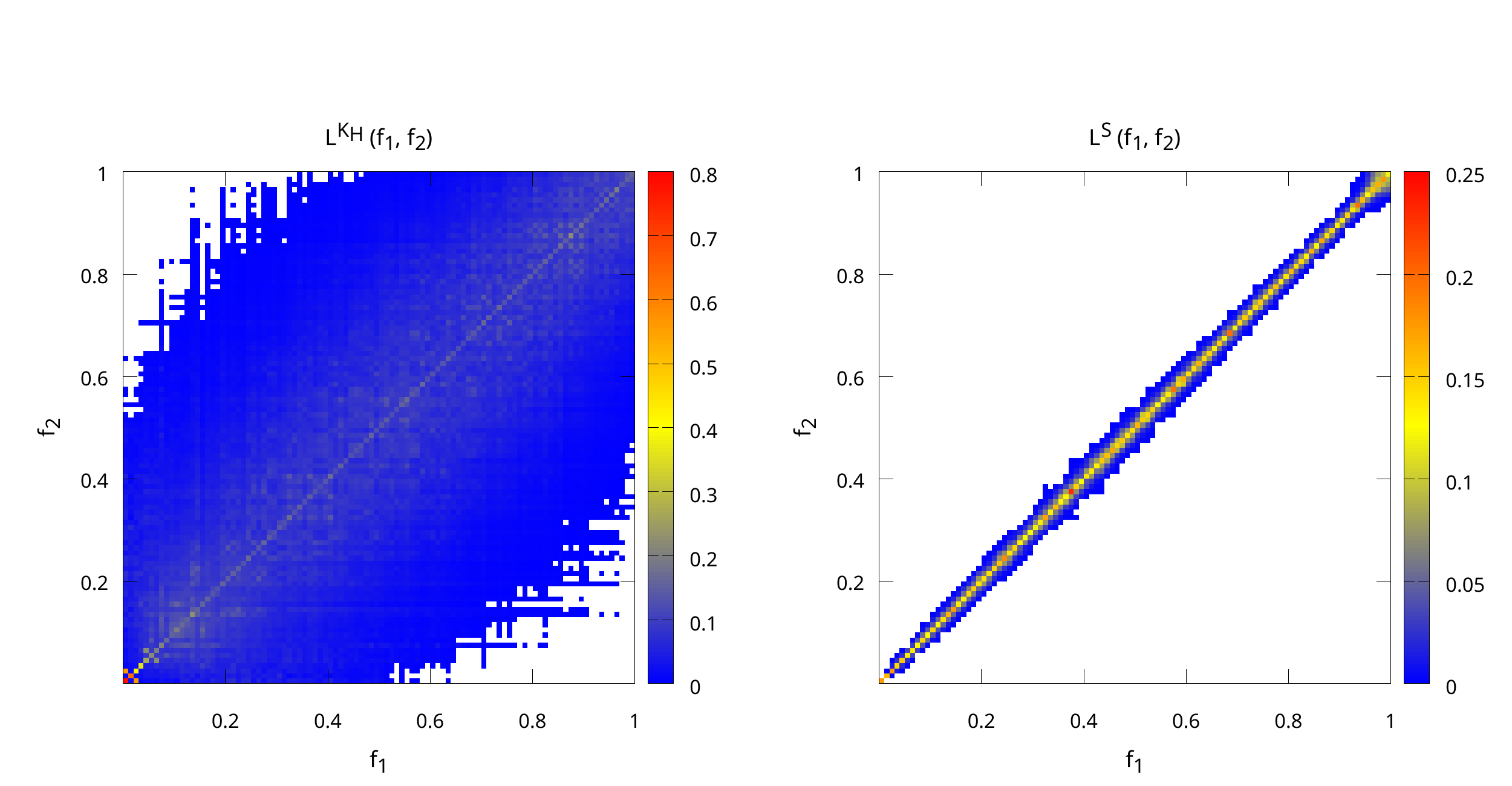}}
\caption{a) 2d plot of the overlap integral $L^{(K_H)}(f_1,f_2)$ between the probability distribution of the Kendal tau distance $K_H$ obtained for different fractions of the randomised candidates $f_1$ and $f_2$. The $L^{(K_H)}(f_1,f_2)$ given in (\ref{eq:ov_int_K_H}) is colour coded, white cells correspond to zero overlap. b) The same 2d plot as in a) for the similarity measure $S$. The $L^{(S)}(f_1,f_2)$ given in (\ref{eq:ov_int_S}) is colour coded, white cells correspond to zero overlap.}
\label{fig:permut_overlaps}
\end{figure}

Furthermore, let us suppose that we would like to guess the fraction of randomised candidates based on the distance value or similarity value in case of e.g., the general randomisation scheme, where candidates are swapped absolutely at random. Due to the large variance in the distance $K_H$, we cannot make a precise guess about $f$, as the range of $f$ values compatible with a given distance is very broad. In contrast, we can make a pretty good guess based on $S$, as the range of $f$ values compatible with a given similarity value is rather narrow. To investigate this aspect of the different comparison methods in more details, we measured the empirical probability distributions of both $K_H$ and $S$ at different fractions of the randomised candidates $f$ when the order of the candidates to be swapped was chosen at random. I.e., at a given $f$, the density $p_f(K_H)$ gives the probability that the Kendal tau distance between the original- and the randomised partial order is $K_H$, and similarly, $p_f(S)$ is corresponding to the probability that the similarity at $f$ is given by $S$. Next we compared the probability distributions for a given measure at different $f$ values by calculating the overlap integrals as
\begin{eqnarray}
L^{(K_H)}(f_1,f_2)&=&\int p_{f_1}(K_H)p_{f_2}(K_H)dK_H, \label{eq:ov_int_K_H}\\
L^{(S)}(f_1,f_2)&=&\int p_{f_1}(S)p_{f_2}(S)dS. \label{eq:ov_int_S}
\end{eqnarray}
Both $L^{(K_H)}(f_1,f_2)$ and $L^{(S)}(f_1,f_2)$ are symmetric by construction. However, according to Fig.\ref{fig:permut_overlaps}., when plotted in a 2d plot, the region where $L^{(K_H)}(f_1,f_2)$ is significantly larger than zero is far more wide compared to the same region for $L^{(S)}(f_1,f_2)$. Thus, the overlap between the probability distributions of $K_H$ at a pair of $f$ values relatively far apart is still large, whereas in contrast the overlap integral between the probability distributions of $S$ quickly becomes zero if the compared $f$ values are not close enough. 

\section*{References}

\bibliography{part_art}

\end{document}